\documentclass[12pt]{article}
\usepackage{varioref}
\usepackage{natbib}
\usepackage{subfigure}
\usepackage{xr-hyper}
\usepackage{booktabs}
\usepackage{multirow}
\usepackage{footnote}
\usepackage{float}
\usepackage{comment}
\usepackage{bm}
\usepackage{amsmath}
\usepackage{amsfonts}
\usepackage{titlesec}
\usepackage{graphicx}
\usepackage[normalem]{ulem}
\usepackage{xcolor}
\usepackage{booktabs}
\usepackage{tabularx}

\usepackage[colorlinks=true,linkcolor=red,citecolor=blue]{hyperref}
\usepackage{bbm}

\newtheorem{theorem}{Theorem}

\def\bSig\mathbf{\Sigma}

\newcommand{\be}{\begin{eqnarray}}
\newcommand{\ee}{\end{eqnarray}}
\newcommand{\bee}{\begin{eqnarray*}}
\newcommand{\eee}{\end{eqnarray*}}
\newcommand{\bi}{\begin{enumerate}}
\newcommand{\ei}{\end{enumerate}}

\newcommand{\one}{\mathbbm{1}}

\newcommand{\bfbeta}{\bm{\beta}}

\newcommand{\blinding}[2]{#1}   

\def\bSig\mathbf{\Sigma}

\begin{document}

\begin{center}
\vspace*{-2.5cm}

{\Large Power analysis for cluster randomized trials with continuous co-primary endpoints}

\medskip

\blinding{
Siyun Yang \quad \quad Mirjam Moerbeek \quad \quad Monica Taljaard \quad \quad Fan Li
\footnote{Siyun Yang is a PhD student at Department of Biostatistics and Bioinformatics, Duke University School of Medicine, Durham, NC, USA; Mirjam Moerbeek is an associate professor at Department of Methodology and Statistics, Utrecht University, Utrecht, The Netherlands; Monica Taljaard is a professor at
 School of Epidemiology and Public Health, University of Ottawa, Ottawa, Ontario, Canada;  Fan Li is an assistant professor at Department of Biostatistics, Yale School of Public Health, New Haven, USA (fan.f.li@yale.edu). }

}{}

\end{center}

\date{}

{\centerline{ABSTRACT}

\noindent Pragmatic trials evaluating health care interventions often adopt cluster randomization due to scientific or logistical considerations. Systematic reviews have shown that co-primary endpoints are not uncommon in pragmatic trials but are seldom recognized in sample size or power calculations. While methods for power analysis based on $K$ ($K\geq 2$) binary co-primary endpoints are available for cluster randomized trials (CRTs), to our knowledge, methods for continuous co-primary endpoints are not yet available. Assuming a multivariate linear mixed model that accounts for multiple types of intraclass correlation coefficients among the observations in each cluster, we derive the closed-form joint distribution of $K$ treatment effect estimators to facilitate sample size and power determination with different types of null hypotheses under equal cluster sizes. We characterize the relationship between the power of each test and different types of correlation parameters. We further relax the equal cluster size assumption and approximate the joint distribution of the $K$ treatment effect estimators through the mean and coefficient of variation of cluster sizes. Our simulation studies with a finite number of clusters indicate that the predicted power by our method agrees well with the empirical power, when the parameters in the multivariate linear mixed model are estimated via the expectation-maximization algorithm. An application to a real CRT is presented to illustrate the proposed method.

\vspace*{0.3cm}
}


\section{Introduction}
\label{s:intro}
The importance of pragmatic trials is increasingly being recognized by patients, clinicians, and health care system stakeholders. Pragmatic trials are distinct from explanatory trials in that they are designed to mimic real-world practice so as to directly inform clinical decision-making \citep{loudon2015precis}. To reflect priorities of multiple stakeholders, pragmatic trials commonly select co-primary endpoints (also known as multiple primary outcomes), for example, to demonstrate effectiveness on both clinical and patient-reported outcomes \citep{Nevins2021}.  In other circumstances, multiple sub-scales from a questionnaire-based scale may be of interest and considered equally important, or an outcome may be assessed on both a patient and their caregiver and analyzed as a multivariate response to account for the mutual influences of the patient and their caregiver on the response to an intervention. Sample size considerations for co-primary endpoints have been previously investigated for individually randomized trials with a recommendation to account for the intra-subject correlation between the co-primary endpoints \citep{Micheaux2014power}. However, methods for designing cluster randomized trials (CRTs) with co-primary endpoints are sparse.

In a CRT, participants are randomized in intact groups, while outcomes are observed on the individual participants \citep{murray1998design}. Cluster randomization is commonly used in pragmatic trials evaluating health care interventions, for example, because the intervention must necessarily be delivered at the cluster level, to avoid contamination, or for logistical reasons. A notable feature of CRTs is that the observations from multiple subjects belonging to the same cluster are no longer independent, and therefore the design and analysis of CRTs necessitate accounting for the intraclass correlation coefficient (ICC). There has been an extensive literature addressing sample size and power calculation procedures for CRTs with a single primary endpoint \citep{rutterford2015methods,turner2017review1a}. In the presence of co-primary endpoints however, the endpoints themselves are usually correlated for the same subject, requiring adjustment for both the intra-subject and inter-subject ICCs during the design phase. For binary co-primary endpoints, \citet{li2020power} developed a power analysis approach based on generalized estimating equations (GEE) with a working independence correlation structure. However, the working independence assumption is often not fully efficient and can lead to a larger sample size than necessary in CRTs even with a single endpoint \citep{li2021sample}. Furthermore, to date, there has been no work investigating power analysis for CRTs with continuous co-primary endpoints. Continuous co-primary endpoints are common in some clinical areas. For example, as found in a review of pragmatic trials in Alzheimer's disease and related dementias, 10 of the 17 trials (59\%) with co-primary outcomes were CRTs \citep{Taljaard2021}, and 9 of these 10 CRTs had continuous co-primary endpoints. Moreover, a review of pragmatic trials across a broad range of clinical areas \citep{Vanderhout2022patient} found that 56 of 415 trials (13.5\%) reported multiple co-primary endpoints, and within that set, 22 out of 152 CRTs (14.5\%) reported multiple co-primary endpoints and 7 had continuous co-primary endpoints (personal communication). These proportions are likely to understate the prevalence of co-primary endpoints, as they represent current practices as opposed to ideal practices when new methods become available.

In this article, we fill the methodological gap by developing an analytical approach for power analysis of CRTs with continuous co-primary endpoints. Our development is based on a multivariate linear mixed model (MLMM), which accounts for the multiple types of correlation parameters in the estimation of treatment effects and therefore has potential to lead to a smaller sample size compared to separate, endpoint-specific analyses. Second, our approach with the MLMM allows for unstructured variance component matrices for random intercepts and random errors, so that in principle, our approach does not require the inter-subject ICCs to be the same across each endpoint. On the contrary, assuming common endpoint-specific ICCs may not always be appropriate, for example, when the selected co-primary endpoints reflect clinical and patient reported measures, or when the co-primary endpoints are measured on patient-caregiver dyads. Finally, we derive the approximate joint distribution of the multivariate test statistic allowing for an unbalanced design with variable cluster sizes. While the impact of variable cluster sizes on sample size calculation for CRTs with a single endpoint has been previously studied \citep{van2007relative}, the impact of variable cluster sizes in CRTs with co-primary endpoints remains unclear. Through analytical derivation and numerical illustration, we show that the efficiency loss due to cluster size variability can be mitigated to some extent by accounting for the co-primary endpoint, thus providing  new motivation for the recommendation to account for co-primary endpoints through a MLMM in both the design and analysis phases of a CRT.

\section{Multivariate Linear Mixed Model}
\label{s:method}
Consider a parallel CRT with $n$ clusters randomly assigned to either control or treatment condition. Suppose continuous co-primary endpoints are measured for each subject, and we define $y_{ijk}$ as the $k$th $(k=1,\dots,K)$ continuous endpoint for the $j$th $(j = 1,\ldots, m_i)$ subject in the $i$th $(i = 1,\ldots, n)$ cluster. Let $\bm{y}_{ij}=(y_{ij1},\ldots,y_{ijK})^T$ denote the collection of all $K$ endpoints for each subject, which is modelled by a MLMM as 
\begin{equation} \label{eq:model1}
 \bm y_{ij}= 
  \left(\begin{array}{@{}c@{}}
                        \gamma_1 \\
                        \vdots \\
                        \gamma_K
  \end{array}\right)+
   \left(\begin{array}{@{}c@{}}
                        \beta_{1} \\
                        \vdots \\
                        \beta_{K}
  \end{array}\right)z_i+
  \left(\begin{array}{@{}c@{}}
                        \phi_{i1}\\
                        \vdots \\
                       \phi_{iK}
  \end{array}\right)+
  \left(\begin{array}{@{}c@{}}
                        e_{ij1} \\
                        \vdots \\
                        e_{ijK}
  \end{array}\right),
\end{equation}
where $z_i$ is the cluster-level treatment indicator with $z_i=1$ for the treatment condition. In model \eqref{eq:model1}, $\gamma_k$ represents the mean of the $k$th endpoint under the control condition, $\beta_k$ represents the average treatment effect for the $k$th endpoint, $\bm{\phi}_i=(\phi_{i1},\dots,\phi_{iK})^T$ is the vector of random intercepts for cluster $i$ across all $K$ endpoints and is assumed to follow $\mathcal{N}(\bm{0}_{K\times 1},\bm{\Sigma_\phi})$, and $\bm{e}_{ij}=(e_{ij1},\dots,e_{ijK})^T$ is the vector of random errors for each subject and follows $\mathcal{N}(\bm{0}_{K\times 1},\bm{\Sigma_e})$. For identifiability, we assume independence between $\bm{\phi}_i$ and $\bm{e}_{ij}$, but do not place further restrictions on $\bm{\Sigma_\phi}$ and $\bm{\Sigma_e}$ other than requiring them to be positive definite. We denote each diagonal element of $\bm{\Sigma_\phi}$ and $\bm{\Sigma_e}$ as $\sigma^2_{\phi k}$ and $ \sigma^2_{ek}$, and off-diagonal element as $\sigma_{\phi kk^{\prime}}$ and $ \sigma_{ekk^\prime}$, thus the marginal variance of each endpoint is $\sigma_{yk}^2=\sigma_{\phi k}^2+\sigma_{e k}^2$, and may vary across $k$. 

With the $K(K+1)/2$ variance component parameters in $\bm{\Sigma_\phi}$ and $K(K+1)/2$ variance component parameters in $\bm{\Sigma_e}$, Table \ref{tb:definition} summarizes the multiple types of ICCs among the endpoints implied from the above MLMM. Specifically, we define (1) $\rho_0^{k}=\text{corr}(y_{ijk},y_{ij^{\prime}k}|z_i)=\sigma_{\phi k}^2/(\sigma_{\phi k}^2+\sigma_{e k}^2) $, representing the inter-subject correlation of the same endpoint, or the endpoint-specific ICC; (2) $\rho_1^{kk^\prime}=\text{corr}(y_{ijk},y_{ij^{\prime} k^{\prime}}|z_i)={\sigma_{\phi kk^\prime}}\Big/\left(\sqrt{\sigma_{\phi k}^2+\sigma_{e k}^2}\sqrt{\sigma_{\phi k^\prime}^2+\sigma_{e k^\prime}^2}\right)$, representing the inter-subject correlation of two outcomes corresponding to two different endpoints $k$ and $k^\prime$, or equivalently the inter-subject between-endpoint ICC; and (3) $\rho_2^{kk^{\prime}}=\text{corr}(y_{ijk},y_{ijk^\prime}|z_i)=\left(\sigma_{\phi kk^\prime}+\sigma_{ekk^\prime}\right)\Big/\left(\sqrt{\sigma_{\phi k}^2+\sigma_{e k}^2}\sqrt{\sigma_{\phi k^\prime}^2+\sigma_{e k^\prime}^2}\right)$, representing the intra-subject between-endpoint ICC, or abbreviated by intra-subject ICC. By the MLMM assumption, we have symmetry such that $\rho_1^{kk^\prime}=\rho_1^{k^\prime k}$, $\rho_2^{kk^{\prime}}=\rho_2^{k^{\prime}k}$ and degeneracy such that $\rho_1^{kk}=\rho_0^k$, $\rho_2^{kk}=1$, $\forall~k,k^\prime$. Define $\bm{\rho}_0=\{\rho_0^k,k=1,\ldots,K\}$, $\bm{\rho}_1=\{\rho_1^{kk^\prime},k\leq k^\prime,k,k^\prime=1,\ldots,K\}$ and $\bm{\rho}_2=\{\rho_2^{kk^\prime},k\leq k^\prime,k,k^\prime=1,\ldots,K\}$. Therefore, $\bm{\rho}_0$, $\bm{\rho}_1$ and $\bm{\rho}_2$ are of size $K$, $K(K-1)/2$ and $K(K-1)/2$, respectively. In addition, there is a one-to-one mapping between the variance component matrices $\{\bm{\Sigma_\phi}$, $\bm{\Sigma_e}\}$ and the set of correlations and marginal variances $\{\bm{\rho}_0$, $\bm{\rho}_1$, $\bm{\rho}_2,\bm{\sigma_y}^2\}$, where $\bm{\sigma_y}^2=\{\sigma_{yk}^2,k=1,\ldots,K\}$. 
Although not required for our methodology, a parsimonious parameterization that does not distinguish the ICCs by different endpoints is $\rho_0^k=\rho_0$, $\rho_1^{kk^\prime}=\rho_1$ and $\rho_2^{kk^\prime}=\rho_2$, $\forall~k,k^\prime$ \citep{li2020power}. This specification engenders the \emph{block exchangeable correlation model} for $\bm{y}_{ij}$, which has been previously proposed for designing longitudinal CRTs \citep{li2018samplesize}.

\begin{table}[htbp]
\caption{Definition and interpretations of multiple ICCs in cluster randomized trials with co-primary endpoints from the multivariate linear mixed model: endpoint-specific ICC ($\rho_0^{k}$), inter-subject between-endpoint ICC ($\rho_1^{kk^\prime}$), and intra-subject ICC ($\rho_2^{k}$) for $k,k^\prime=1,\ldots,K$.}\label{tb:definition}
\vspace{0.2cm}
\begin{tabularx}{1\textwidth}{>{\arraybackslash}m{4.3cm}>{\centering\arraybackslash}m{4.3cm}>{\arraybackslash}m{6.9cm}}
\toprule
Definition  & Expression  &Interpretation \\
\midrule
$\rho_0^{k}=\text{corr}(y_{ijk},y_{ij^{\prime}k}|z_i)$ & $\displaystyle\frac{\sigma_{\phi k}^2}{\sigma_{\phi k}^2+\sigma_{e k}^2}$ &  The intraclass correlation parameter between two outcomes from subject $j$ and $j^\prime$ but corresponding to the same endpoint $k$, or \emph{endpoint-specific ICC}.\\
\midrule
$\rho_1^{kk^\prime}=\text{corr}(y_{ijk},y_{ij^{\prime} k^{\prime}}|z_i)$ & $\displaystyle\frac{\sigma_{\phi kk^\prime}}{\sqrt{\sigma_{\phi k}^2+\sigma_{e k}^2}\sqrt{\sigma_{\phi k^\prime}^2+\sigma_{e k^\prime}^2}}$ & The intraclass correlation parameter between two outcomes from subject $j$ and $j^\prime$ and corresponding to two different endpoint $k$ and $k^\prime$, or \emph{inter-subject between-endpoint ICC}.\\
\midrule
$\rho_2^{k}=\text{corr}(y_{ijk},y_{ijk^\prime}|z_i)$ & $\displaystyle\frac{\sigma_{\phi kk^\prime}+\sigma_{ekk^\prime}}{\sqrt{\sigma_{\phi k}^2+\sigma_{e k}^2}\sqrt{\sigma_{\phi k^\prime}^2+\sigma_{e k^\prime}^2}}$ & The intraclass correlation parameter between two outcomes from the same subject $j$ but corresponding to two different endpoint $k$ and $k^\prime$, or \emph{intra-subject (between-endpoint) ICC}.\\ 
\bottomrule
\end{tabularx}
\end{table}

While our primary focus is power analysis of CRTs with co-primary endpoints based on MLMM \eqref{eq:model1}, we also include details for estimating the MLMM parameters during the analytical stage in {Web Appendix A}. We adopt the expectation–maximization (EM) algorithm by treating the random intercepts as missing variables and outline the iterative approach for estimating both the treatment effect and variance component matrices. The associated standard errors are then obtained from numerically differentiating the log-likelihood function evaluated at the maximum likelihood estimators. In Section \ref{s:simu}, we demonstrate via simulations that the EM approach can provide good control of type I error rate and precise empirical power compared to formula predictions.


\section{Power Analysis of CRTs with Continuous Co-Primary Endpoints}\label{sec:samplesize}
\subsection{Joint distribution of $K$ treatment effect estimators}\label{sec:jointdis}

We derive the analytical form of the joint distribution of the $K$ Wald test statistics for the $K$ treatment effect parameters $\bfbeta=(\beta_1,\ldots,\beta_K)^T$ in the MLMM. To begin with, we assume the cluster sizes are all equal such that $m_i=m$. We define $\overline{z} = \mathbb{E}(z_i)$ as the allocation probability at the cluster level and re-parameterize the MLMM such that
\begin{align} \label{eq:model2}
\bm y_{ij}
   =\left(\begin{array}{@{}c@{}}
                        \widetilde{\gamma}_1 \\
                        \vdots \\
                        \widetilde{\gamma}_K
  \end{array}\right)+
   \left(\begin{array}{@{}c@{}}
                        \beta_{1} \\
                        \vdots \\
                        \beta_{K}
  \end{array}\right)(z_i-\overline{z})+
   \left(\begin{array}{@{}c@{}}
                        \phi_{i1}\\
                        \vdots \\
                       \phi_{iK}
  \end{array}\right)+
  \left(\begin{array}{@{}c@{}}
                        e_{ij1} \\
                        \vdots \\
                        e_{ijK}
  \end{array}\right),
\end{align}
where the updated intercept for each endpoint is $\widetilde{\gamma}_k=\gamma_k+\beta_k \overline{z}$, while the treatment effect parameter for each endpoint remains unchanged. 
Recall that the total marginal variance of $y_{ijk}$ conditional on $z_i$ is $\text{var}(y_{ijk}|z_i)=\sigma^2_{\phi k}+\sigma^2_{ek}$, and the covariance for any pair of endpoints in the same cluster is 
\begin{equation*}
\text{cov}(y_{ijk},y_{ij'k^\prime}|z_i)=  \sigma^2_{\phi k}\one\{j\ne j', k = k^\prime\} + \sigma_{\phi kk^\prime}\one\{j\ne j', k \ne k^\prime\}+(\sigma_{\phi kk^\prime}+\sigma_{e kk^\prime})\one\{j= j', k \ne k^\prime\},
\end{equation*}
where $\one(\bullet)$ is the indicator function. In matrix notation, let 
$\bm{y}_i=(\bm {y}_{i1}^T,\ldots,\bm {y}_{i m}^T)^T$ denote the vector of all co-primary endpoints in cluster $i$, then the MLMM implies the covariance matrix $\bm{V_i}=\text{cov}(\bm{y}_i| z_i )=\bm{I}_{m} \otimes \bm{\Sigma_e}+\bm{J}_{m}\otimes \bm{\Sigma_{\phi}}$ ,where $\bm{I}_{m}$ is the $m\times m$ identity matrix, $\bm{J}_{m}$ is the $m\times m$  matrix of ones. Furthermore, we write the design matrix for each cluster based on model \eqref{eq:model2} as $\bm{W}_i=\bm{1}_m\otimes (\bm{I}_K, \bm{I}_K(z_i-\overline{z}))$, and $\bm{W}_{ij}=(\bm{I}_K, \bm{I}_K(z_i-\overline{z}))$ as the design matrix for each subject. 
Then the best linear unbiased estimator of $\bm\theta=( \widetilde{\gamma}_1,\ldots,\widetilde{\gamma}_K,\beta_1,\ldots,\beta_K)^T$ is given by the Feasible Generalized Least Square (FGLS) estimator, denoted by $\widehat{\bm{\theta}}=\left(\sum_{i=1}^{n} \bm{W}_i^T \bm{V}_i^{-1}\bm{W}_i\right)^{-1}\left(\sum_{i=1}^{n} \bm{W}_i^T \bm{V}_i^{-1}\bm{y}_i\right)$, whose
large-sample variance is given by $\text{var}(\widehat{\bm{\theta}})=\bm{U}_n^{-1}$, with $\bm{U}_n=\sum_{i=1}^{n} \bm{W}_i^T \bm{V}_i^{-1}\bm{W}_i$. We derive an explicit form of $\text{var}(\widehat{\bm{\theta}})$ to facilitate analytical power analysis in the design stage. Specifically, we use the results in \citet{leiva2007linear} to obtain the inverse of $\bm{V_i}$ as
\begin{align}
\bm{V}_i^{-1}=\bm{I}_{m}\otimes \bm{\Sigma_e}^{-1}+
\bm{J}_{m}\otimes\frac{1}{m}
\left\{\left(\bm{\Sigma_e}+m\bm{\Sigma_\phi}\right)^{-1}-\bm{\Sigma_e}^{-1}\right\}\label{eq:vinv}.
\end{align}
{The explicit inverse \eqref{eq:vinv} facilitates the simplification of $\bm{U}_n$ and leads to the following result.
\begin{theorem}\label{thm0}
Under the MLMM \eqref{eq:model2} and assuming equal cluster sizes such that $m_i=m$, the FGLS estimator for the vector of treatment effect estimators\\ $\widehat{\bm{\beta}}=(\widehat{\beta}_1,\ldots,\widehat{\beta}_K)^T=\left\{\sum_{i=1}^n (z_i-\overline{z})^2\right\}^{-1}\left\{\sum_{i=1}^n\sum_{j=1}^m m^{-1}(z_i-\overline{z})\bm{y}_{ij}\right\}$ and is free of any ICCs. Furthermore, the lower right $K\times K$ block of $\bm{\Omega_\beta}$, or the asymptotic variance of the scaled FGLS estimator $\sqrt{n}(\widehat{\bm{\beta}}-\bm{\beta})$, has a simple form
\be\label{eq: omegab}
\bm{\Omega_{\beta}}=\frac{1}{m\sigma_z^2}\left(\bm{\Sigma_e}+m\bm{\Sigma_{\phi}} \right).
\ee
Based on the one-to-one mappings from the variance component matrices and the three types of ICCs, the covariance parameters in the joint distribution of $\widehat{\bm{\beta}}$ are equivalently written as
\begin{align}
\omega^2_{k}&=n\text{var}(\widehat\beta_k)=\frac{\left(\sigma_{\phi k}^2+\sigma_{ek}^2\right)\left\{1+(m-1)\rho_0^{k}\right\}}{m\sigma_z^2}\label{eq:var} \\
\omega_{kk^\prime}&=n\text{cov}(\widehat\beta_k,\widehat\beta_{k^\prime})=\frac{\sqrt{\sigma_{\phi k}^2+\sigma_{e k}^2}\sqrt{\sigma_{\phi k^\prime}^2+\sigma_{e k^\prime}^2}\left\{ \rho_2^{kk^\prime}+(m-1)\rho_1^{kk^\prime}\right\}}{m\sigma_z^2} \label{eq:cov},    
\end{align}
for $k =1,\dots,K$, and $k^\prime\ne k$. One can further set $\sigma_z^2=1/4$ under equal treatment allocation.
\end{theorem}

 The proof of Theorem \ref{thm0} is found in Web Appendix B. Several comments are in order based on equations \eqref{eq:var} and \eqref{eq:cov}. First, under the assumption of equal cluster sizes ($m_i=m$), the variance of the treatment effect estimator corresponding to each endpoint, $\omega_k^2$, based on the MLMM, is identical to that obtained by analyzing each endpoint via a separate linear mixed model (LMM). In particular, the variance inflation factor (VIF) for estimating $\beta_k$ in a CRT relative to an individually randomized trial (IRT) equals the usual VIF, $1+(m-1)\rho_0^{k}$, which is an increasing function of $m$ and the endpoint-specific ICC (but not other types of ICCs). If the interest lies in testing $H_0:\beta_k=0$ for one specific endpoint $k$, then the conventional power analysis approach developed for CRTs with a single endpoint \citep{murray1998design} can be directly used even if a MLMM is considered in the primary analysis. Second, the MLMM allows for objective comparisons between the treatment effects across different endpoints, but the power of such comparisons can depend on the covariance parameter, $\omega_{kk^\prime}$, which is an increasing function of the marginal variance of endpoints $k$, $k^\prime$, the inter-subject between-endpoint ICC, $\rho_{1}^{kk^\prime}$, and the intra-subject ICC, $\rho_2^{kk^\prime}$. Expression \eqref{eq:cov} further suggests the covariance inflation factor (cVIF) in a CRT relative to an IRT with co-primary endpoints is given by
$\text{cVIF}=1+(m-1)\left(\rho_1^{kk^\prime}/\rho_2^{kk^\prime}\right)$, which is an increasing function of cluster size $m$ as well as the ratio between the inter-subject between-endpoint ICC and intra-subject ICC. 
In particular, a larger cluster size increases the magnitude of each element in the covariance matrix for the $K$ treatment effect estimators, but at a differential rate for the variance element (rate of increase is $\rho_0^{k}$) and the covariance element (rate of increase is $\rho_1^{kk^\prime}/\rho_2^{kk^\prime}$). The explicit characterization of the joint distribution for $\sqrt{n}(\widehat{\bm{\beta}}-\bm{\beta})$ based on the MLMM allows us to develop an analytical sample size procedure for testing any general linear hypothesis concerning the treatment effects. 

\subsection{Power analysis for testing general linear hypotheses}
\label{s:testequality}
The characterization of $\bm{\Omega}_{\bm{\beta}}$ provides an analytical approach to quantify the power of any general linear hypothesis test concerning the treatment effect parameters $\bm{\beta}$ in the MLMM \eqref{eq:model1}. Specifically, a testable general linear hypothesis of interest in CRTs can be written as $H_0:\bm{ L\beta}=\bm{0}$, vs. $\bm{ L\beta} \ne \bm{0}$, where $\bm L$ is a $S\times K$ ($S\leq K$) contrast matrix whose rows represent linearly independent hypotheses concerning the treatment effect parameter $\bm\beta$. A commonly used test statistic for $H_0$ is the $F$-statistic \citep{roy2007sample}, 
\be \label{eq:Fstats}
F^*=\frac{n(\bm{L}\widehat{\bm{\beta}})^T(\bm{L}\widehat{\bm{\Omega}}_{\bfbeta}\bm{L}^T)^{-1}(\bm{L}\widehat{\bfbeta})}{S},
\ee
where $\widehat{\bm{\Omega}}_{\bm{\beta}}$ is the estimated variance-covariance matrix for the treatment effect estimator $\widehat{\bm{\beta}}$. Under the null, $F^*$ approximately follows a central $F$-distribution with numerator and denominator degrees of freedoms $(S,\nu)$. Under the alternative, $F^*$ approximately follows a non-central $F$-distribution with non-centrality parameter $\tau$, and degree of freedom $(S,\nu)$, where $\nu$, for example, can be specified as $n-S-K$. Of note, in a CRT with a single endpoint such that $S=K=1$, this degree of freedom coincides with the \emph{between-within degree of freedom} and has been previously demonstrated to have adequate control of type I error rate in CRTs with a small number of clusters \citep{li2017evaluation}. The non-centrality parameter can be approximated by $\widehat{\tau}=n(\bm{ L}\widehat{\bfbeta})^T(\bm{L}\widehat{\bm\Omega}_{\bfbeta}\bm{L}^T)^{-1}(\bm{ L}\widehat{\bfbeta})$, and serves as a basis for power analysis. For pre-specified type I error rate $\alpha$, the power under $H_1:\bm{ L\beta}=\bm{\delta}\neq \bm{0}$ based on \eqref{eq:Fstats} is
\begin{equation} \label{eq: power1}
1-\lambda = \int_{F_{1-\alpha}(S,n-S-K)}^{\infty} f(x; \tau, S,n-S-K)dx,
\end{equation}
where $\lambda$ is the type II error rate, $F_{1-\alpha}(S,n-S-K)$ is the critical value of the central $F(S, n-S-K)$ distribution, and $f(x;\tau, S,n-S-K)$ is the probability density function of the non-central $F(\tau,S,n-S-K)$ distribution with non-centrality parameter $\tau=n\bm{\delta}^T(\bm{L}{\bm\Omega}_{\bfbeta}\bm{L}^T)^{-1}\bm{\delta}$. Alternatively, equation \eqref{eq: power1} can be numerically solved to determine the required number of clusters or cluster size to achieve a desired level of power. With $K$ co-primary endpoints in a CRT, we explore the relationship between the three types of ICC parameters and the power of typical hypothesis tests. First, we focus on the \emph{omnibus test} for detecting any departure from the global null $H_0:\beta_k=0$ $\forall~k$, corresponding to a contrast matrix $\bm{L}=\bm{I}_K$ (i.e., $S=K$). For this test, rejecting $H_0$ indicates that the treatment has a statistically significant effect on at least one of the endpoints. In Web Appendix C, we show that with all other design parameters fixed, a larger value of the endpoint-specific ICC, $\rho_0^{k}~\forall~k$, is always associated with a smaller power of the omnibus test, namely, a larger required sample size, while the relationship between $\rho_1^{kk^\prime}$, $\rho_2^{kk^\prime}$ $\forall~k\neq k^\prime$ and the power of the omnibus test is generally indeterminate. However, under a simpler parameterization with the block exchangeable correlation structure, we further obtain the following result.
\begin{theorem}\label{thm1}
(Omnibus test) Under the parsimonious block exchangeable correlation structure such that $\rho_0^{k}=\rho_0$, $\rho_1^{kk^\prime}=\rho_1$, $\rho_2^{kk^\prime}=\rho_2$ $\forall~k,k^\prime$, and assuming equal standardized effect sizes such that ${\beta_k}/\sigma_{yk}={\beta_{k^\prime}}/\sigma_{yk^\prime}$ $\forall~k\neq k^\prime$, a larger value of the endpoint-specific ICC, $\rho_0$, and larger values of between-endpoint ICCs, $\rho_1$ or $\rho_2$, are always associated with a smaller power of the omnibus test (larger sample size).
\end{theorem}

Because the power of the $F$-test is an increasing function of the non-centrality parameter, the proof of Theorem \ref{thm1} boils down to assessing the monotonicity of $\tau$ as a function of different types of ICCs. Without further assumptions, we show in Web Appendix B that larger values of the endpoint-specific ICCs, $\bm{\rho}_0$, lead to smaller power of the omnibus test, suggesting that $\bm{\rho}_0$ plays a similar role in CRTs with co-primary endpoints as the conventional ICC does in a CRT with a single primary endpoint. Ignoring $\bm{\rho}_0$ in the design stage will necessarily result in sample size being under-estimated. While the relationship between $\bm{\rho}_1$, $\bm{\rho}_2$ and the power is generally unclear, Theorem \ref{thm1} clarifies the role of the between-endpoint ICCs for study power under further restrictions on correlations across endpoints (i.e., assuming the block exchangeable correlation structure and equalizing the standardized effect sizes). That is, ignoring the common between-endpoint ICCs ($\rho_1$ or $\rho_2$) will result in sample size being under-estimated when studying the global hypothesis with the omnibus test.

In Web Appendix B, we additionally explore the \emph{test for treatment effect homogeneity} across $K$ endpoints with $H_0:\beta_k=\beta_{k^\prime}$ $\forall~k\neq k^\prime$, corresponding to a contrast matrix $\bm{L}=(\bm{e}_1-\bm{e}_2,\bm{e}_2-\bm{e}_3,\ldots,\bm{e}_{K-1}-\bm{e}_K)^T$ (i.e., $S=K-1$), where $\bm{e}_k$ is the $K\times 1$ vector with $1$ at the $k$th position and zero elsewhere. Rejecting $H_0$ implies that the treatment effect is different for at least one of the co-primary endpoints compared to the other endpoints. In Web Appendix B, we prove that the endpoint-specific ICC plays a similar role in the power of the test for treatment effect homogeneity just like the omnibus test. However, in contrast to Theorem \ref{thm1}, larger values of between-endpoint ICCs are associated with a larger power of the test for homogeneity under the parsimonious block exchangeable correlation structure (smaller required sample size). We provide a summary of these relationships in Table \ref{tab:relationship}.

\begin{table}[htbp]
  \begin{center}
    \caption{Concise summary of the relationships between ICC parameters and the power of tests. Abbreviation and notation: `General' indicates no restrictive assumption on the correlation structure, `BEX' stands for block exchangeable correlation structure; `INDET' indicates that the relationship is indeterminate; `$\Uparrow$' indicates a monotonically increasing relationship, and `$\Downarrow$' indicates a monotonically decreasing relationship.}
    \label{tab:relationship}
    \begin{tabular}{l|c|c|c|c}
          \toprule
      Test & Assumption & $\bm{\rho_0}$ & $\bm{\rho_1}$ & $\bm{\rho_2}$\\
      \midrule
      \multirow{2}{*}{Omnibus test} & General & $\Downarrow$ & INDET & INDET \\ 
       & BEX, $\{{\beta_k}/\sigma_{yk}={\beta_{k^\prime}}/\sigma_{yk^\prime},~\forall~k\neq k^\prime\}$ & $\Downarrow$ & $\Downarrow$ & $\Downarrow$\\ 
      \midrule
      \multirow{2}{*}{Test for effect homogeneity} & General & $\Downarrow$ & INDET & INDET\\ 
       & BEX, $\{\sigma_{yk}^2=\sigma_{yk^\prime}^2,~\forall~k\neq k^\prime\}$ & $\Downarrow$ & $\Uparrow$ & $\Uparrow$ \\ 
      \midrule
      \multirow{2}{*}{Intersection-union test} & General & $\Downarrow$ & $\Uparrow$ & $\Uparrow$\\ 
       & BEX & $\Downarrow$ & $\Uparrow$ & $\Uparrow$ \\ 
      \bottomrule
    \end{tabular}
  \end{center}
\end{table}

\subsection{Power analysis for simultaneously testing treatment effects across all endpoints}\label{s:testoverall}

With co-primary endpoints, the intersection-union test has also considered to avoid inflation of the type I error rate
\citep{chuang2007challenge,sozu2010sample}. Unlike the omnibus test whose null hypothesis is simple, the intersection-union test focuses on a simple alternative but a composite null such that the test rejects only when the treatment effect is non-zero across all endpoints. We consider testing $H_0:\beta_k=0$ for at least one $k$ against the one-sided alternative $H_1:\beta_k>0$ $\forall~k$. Although we focus on the one-sided alternative, extensions to a class of non-inferiority tests or a two-sided intersection union test are also straightforward with the following characterization of the joint distribution of the test statistics (an example of two-sided intersection-union test can be found in \citet{Tian2021}). For testing $H_0$, we consider the vector of Wald test statistics $\bm{\zeta}=(\zeta_1,\ldots,\zeta_K)^T$, where $\zeta_k=\sqrt{n}\widehat{\beta}_k /\widehat{\omega}_k$, $\widehat{\omega}_k$ is the estimated standard error of the treatment effect estimator from the MLMM \eqref{eq:model1}. Based on the results in Section \ref{sec:jointdis}, because the test statistic is standardized by the standard error, the vector of standardized test statistics $\bm{\zeta}$ asymptotically follows a multivariate normal distribution with mean $\bm{\eta}=(\sqrt{n}\beta_1/\omega_{1},\ldots,\sqrt{n}\beta_K/\omega_{K})^T$ and correlation matrix $\bm{\Phi}$, whose diagonal and off-diagonal elements are given by
\begin{equation}\label{eq:phi}
\phi_{kk^\prime}=\one\{k=k^\prime\}+
\frac{\omega_{kk^\prime}}{\omega_k\omega_{k^\prime}}
\one\{k\ne k^\prime\}.
\end{equation}

Given the total number of clusters $n$, cluster size $m$, as well as the true effect size parameters $\bm{\beta}$, the power function to simultaneously detect the effect for all $K$ endpoints are
\begin{equation}\label{eq: power2}
1-\lambda =\mathbb{P}\left\{\mathcal{R}=\bigcap_{k=1}^K\left(\zeta_k>c_k\right)\mid H_1\right\}= \int_{c_1}^{\infty}\ldots\int_{c_K}^{\infty} f_{\bm W}(w_1,\dots,w_K) d{w_1}\ldots d{w_K},
\end{equation}
where $\mathcal{R}$ denotes the pre-specified rejection region, $\{c_1,\ldots,c_K\}$ are the corresponding endpoint-specific critical values for rejection, and $f_{\bm W}$ is the density function of the Wald test statistics under the alternative. While a typical choice of $f_{\bm W}$ is the multivariate normal distribution with mean $\bm{\eta}$ and covariance matrix $\bm{\Phi}$, a multivariate $t$-distribution with location vector $\bm{\eta}$, shape matrix $\bm{\Phi}$, and degrees of freedom $n-2K$ can account for the uncertainty in estimating the covariance parameters and better control for the type I error rate with a limited number of clusters \citep{li2020power,Tian2021}. We henceforth assume $f_{\bm W}$ to be the multivariate $t$-distribution for design calculations throughout. The specification of critical values $\bm{c}$ can lead to an intersection-union test with different operating characteristics \citep{kordzakhia2010method}. We follow the simple approach suggested by \citet{li2020power} such that $c_1=\dots=c_K= t_{\alpha}(n-2K)$, where $t_{\alpha}(n-2K)$ is the ($1-\alpha$) quantile of the univariate $t$ distribution. This specification of critical values is conservative such that the type I error rate is controlled strictly below $\alpha$ within the composite null space ($H_0:\beta_k=0$ for at least one $k$). In the most extreme case where all but one endpoint correspond to a large treatment effect, the size of the test is exactly $\alpha$. Of note, the performance of this approach can critically depend on the number of clusters and number of endpoints. For example, when the number of clusters is small but several endpoints are being considered, the estimated degrees of freedom $n-2K$ may be very small and therefore, the test may be conservative. While we have presented the power equation in \eqref{eq: power2}, we can obtain the required sample size based on the target power $1-\lambda$ by solving for $m$ or $n$ based on any standard iterative algorithm. Finally, similar to testing the general linear hypotheses in Section \ref{s:testequality}, there exists a monotonic relationship between the power of the intersection-union test and the three types of ICC parameters, without further restrictions on the variance components. 
\begin{theorem}  
(Intersection-union test) With all other design parameters fixed, a larger value of the endpoint-specific ICC, $\rho_0^{k}~\forall~k$, is always associated with a smaller power of the intersection-union test (a larger sample size), whereas a larger value of the between-endpoint ICC, $\rho_1^{kk^\prime}$ or $\rho_2^{kk^\prime}$ $\forall~k \ne k^\prime$, is always associated with a larger power (a smaller sample size).
\label{thm3}
\end{theorem}
The proof of Theorem \ref{thm3} can be found in Web Appendix D. Evidently, a larger endpoint-specific ICC leads to a larger required sample size for the intersection-union test and therefore ignoring the endpoint-specific ICC can result in an under-powered trial. This observation suggests that the endpoint-specific ICC plays the same role in the power of the intersection-union test as the omnibus test (Theorem \ref{thm1}). On the contrary, a larger value of any between-endpoint ICC will lead to a higher power of the intersection-union test. During the design stage, Theorem \ref{thm3} suggests that assuming smaller values for any between-endpoint ICC will increase the required sample size and will therefore be a conservative approach. Finally, with continuous co-primary endpoints, Theorem \ref{thm3} can be considered as the analogue of the result of \citet{li2020power} derived for binary co-primary endpoints, even though we have focused on the analysis with MLMM in contrast to that earlier work using independence GEE. 

\subsection{Generalization to accommodate unequal cluster sizes for power analysis}
\label{s:cv}
To operationalize the power analysis of CRTs with continuous co-primary endpoints in more pragmatic settings, we further develop an approximate variance expression for the treatment effect estimators, $\widehat{\bm{\beta}}$, when the cluster sizes, $m_i$, are variable. To derive a modified variance expression adjusting for unequal cluster sizes,  we assume the cluster sizes come from a common distribution $f(m_i)$ with bounded mean $\overline{m}$ and variance $\sigma_m^2$.} 
Using the Neumann series for matrix inverse up to the second order \citep{beilina2017numerical}, we show in Web Appendix E that the limit of the variance of the scaled FGLS estimator $\sqrt{n}(\widehat{\bm{\beta}}-\bm{\beta})$ takes the form
\begin{align}
\bm{\Omega_{\beta}}
\approx \frac{(\bm{\Sigma_e}+\overline m \bm {\Sigma_{\phi}})}{\overline m\sigma_z^2}\times
\underbrace{\left[\bm{I}_K -\text{CV}^2\left\{\overline m \bm {\Sigma_{\phi}}(\bm{\Sigma_e}+\overline m\bm {\Sigma_{\phi}})^{-1}\bm{\Sigma_e}(\bm{\Sigma_e}+\overline m\bm {\Sigma_{\phi}})^{-1}\right\}\right]^{-1}}_{\text{Correction Matrix~}\bm{\Theta}},\label{eq:vark} 
\end{align}
where we define  $\text{CV}=\sigma_m/m$ as the coefficient of variation for $f(m_i)$, and $\bm{\Theta}$ can be considered as a correction matrix to \eqref{eq: omegab} due to unequal cluster sizes. In Web Appendix E, we also derive a correction matrix using the Neumann series up to the fourth order, which further depends on the skewness and kurtosis of the cluster size distribution $f(m_i)$. In practice, however, the higher-order moments of $f(m_i)$ is often more difficult to elicit in the design stage, which renders the more complex fourth-order approximation less useful. We also show in the ensuing simulation study that the second-order approximation \eqref{eq:vark} already provides an adequate approximation for power analysis. Finally, when the between-endpoint ICCs are all identically zero such that $\rho_1^{kk^\prime}=\rho_2^{kk^\prime}=0$ $\forall~k\neq k^\prime$, the variance of the treatment effect estimator $\beta_k$ based on \eqref{eq:vark} becomes
\be
\omega^2_{k}\approx\frac{\left(\sigma_{\phi k}^2+\sigma_{ek}^2\right)\left\{1+(m-1)\rho_0^{k}\right\}}{m\sigma_z^2}
\times \underbrace{\left\{1-\text{CV}^2\frac{\overline{m}\rho_0^{k}(1-\rho_0^{k})}{\{1+(\overline{m}-1)\rho_0^{k}\}^2}\right\}^{-1}}_{\text{Correction Factor~}\theta_k},
\label{eq:varunequal}
\ee
and the correction factor due to unequal cluster sizes $\theta_k$ reduces to the familiar expression derived in \citet{van2007relative} with a single primary endpoint. In more general cases, the diagonal element in \eqref{eq:vark} is different from \eqref{eq:varunequal} and power analysis for the general linear hypothesis test and the intersection-union test should proceed based on \eqref{eq:vark}.
\begin{figure}[htbp] 
\centering
\includegraphics[scale=0.6]{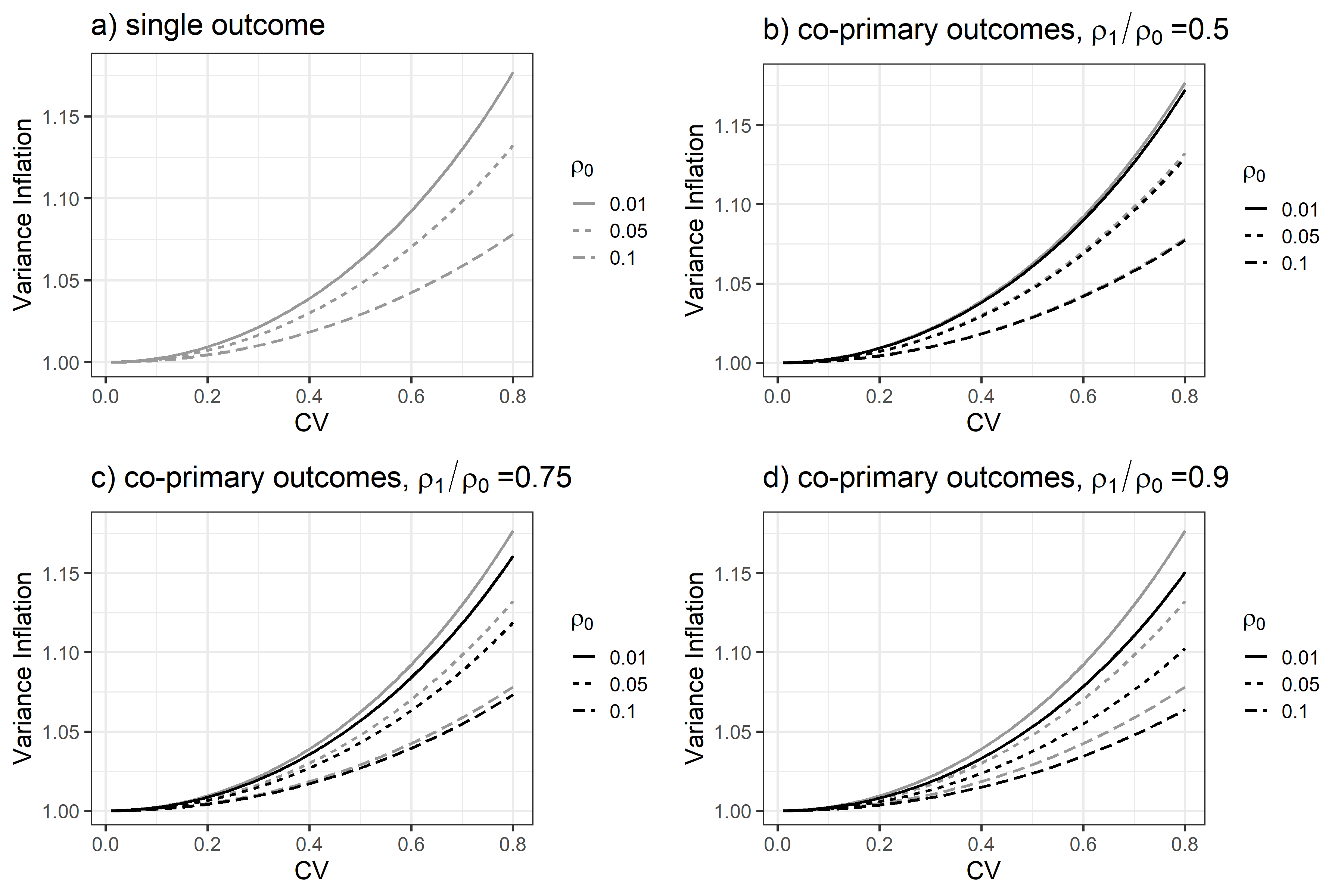}
\caption{Correction factor or variance inflation due to unequal cluster sizes for MLMM and separate LMM analyses of CRTs with co-primary endpoints. (a) Variance inflation for the treatment effect estimator for separate LMM analysis of each endpoint. (b) Variance inflation for the treatment effect estimator for MLMM analysis of two co-primary endpoints when $\rho_1 / \rho_0 =0.5$, $\rho_2=0.2$; (c) Variance inflation for the treatment effect estimator for MLMM analysis of two co-primary endpoints when $\rho_1 / \rho_0 =0.75$, $\rho_2=0.2$; (d) Variance inflation for the treatment effect estimator for MLMM analysis of two co-primary endpoints when $\rho_1 / \rho_0 =0.9$, $\rho_2=0.2$. In (b-d), the gray lines replicate the results in (a) and facilitate efficiency comparisons between MLMM and separate LMM analyses.}
\label{fig:numeri}
\end{figure}

\begin{figure}[htbp]
\centering
\includegraphics[scale=0.6]{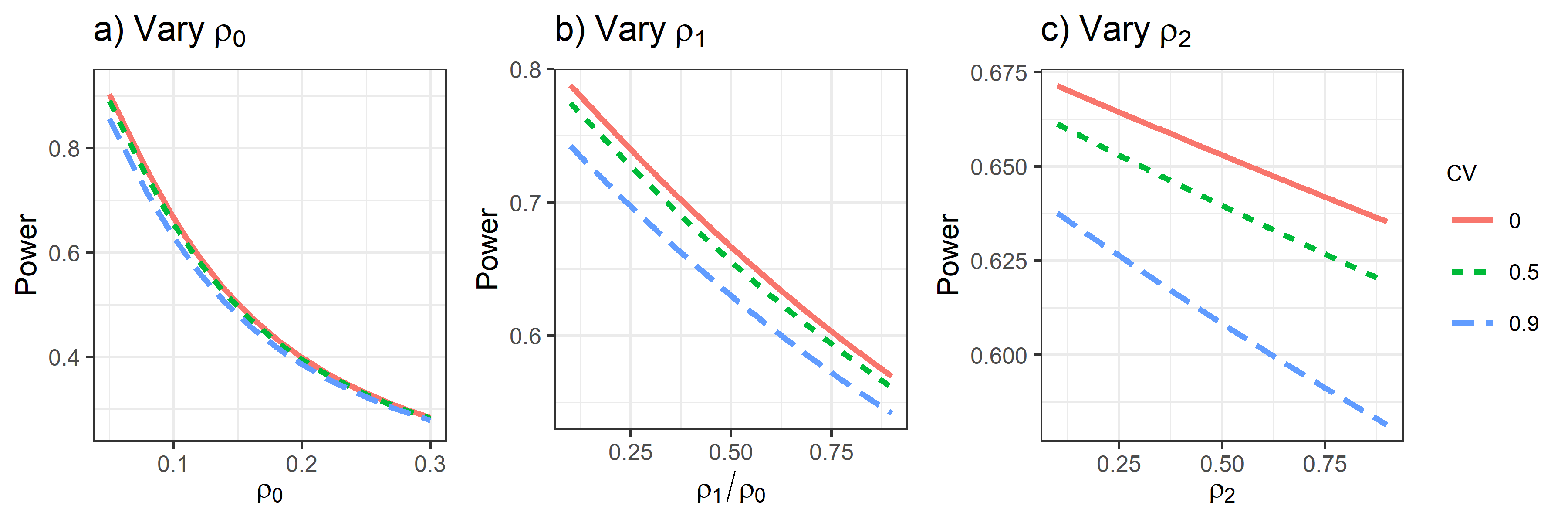}
\caption{Power of the omnibus test with $K=2$ co-primary endpoints as a function of (a) endpoint-specific ICC $\rho_0$, when fixing $\rho_1/\rho_0=0.5$ and $\rho_2=0.2$; (b) inter-subject between-endpoint ICC $\rho_1$\, when fixing $\rho_0=0.1$ and $\rho_2=0.2$; (c) intra-subject ICC $\rho_2$, when fixing $\rho_0=0.1$ and $\rho_1/\rho_0=0.5$. All scenarios assume $n= 30$, $\overline m=60$, $\bfbeta=(0.3,0.3)^T$, $\sigma_{yk}^2=1$ and equal randomization with $\sigma_z^2={1}/{4}$. All figures assume the block exchangeable correlation structure such that $\rho_0^{k}=\rho_0$, $\rho_1^{kk^\prime}=\rho_1$, $\rho_2^{kk^\prime}=\rho_2$ for $k\neq k^\prime\in\{1,2\}$.}\label{fig2:omnibus}
\end{figure}

Under unequal cluster sizes, the multiplicative correction matrix $\bm{\Theta}$ in variance expression \eqref{eq:vark} suggests that the efficiency for estimating the treatment effect for endpoint $k$ can differ when the analysis proceeds with MLMM or when the analysis proceeds with a separate LMM for each endpoint $k$. To illustrate their difference, we numerically compare the efficiency for estimating $\beta_k$ using MLMM and separate LMM. Specifically, we consider a CRT with $K=2$ continuous co-primary endpoints with equal randomization such that $\sigma_z^2=1/4$. We assume mean cluster size $\overline{m}=60$, unit marginal total variance for each endpoint, and a block exchangeable correlation structure such that $\rho_0=\rho_0^{1}=\rho_0^2$. Figure \ref{fig:numeri}(a) presents the values of $\theta_k$ as a function of CV$\in[0,0.8]$ and three different values of endpoint-specific ICC $\rho_0 \in \{0.01, 0.05, 0.1\}$. These values are commonly used in the literature to investigate the impact of unequal cluster sizes on the efficiency of estimating $\beta_k$ from a separate LMM, as in \citet{eldridge2006sample} and \citet{van2007relative}. When $\text{CV}=0$, the cluster sizes are all equal to $60$, and when $\text{CV}=0.8$, the cluster sizes can have substantial variability, ranging from $2$ to $200$, as the density plot in Web Figure 1 demonstrates. Figure \ref{fig:numeri}(b-d) additionally present the values of $\bm{\Theta}_{kk}$ (diagonal values of the correction matrix $\bm{\Theta}$) on the same set of design parameters but with $\rho_1/\rho_0\in \{0.5, 0.75, 0.9\}$ and $\rho_2=0.2$. These values correspond to the impact of unequal cluster sizes on the efficiency of estimating $\beta_k$ from the MLMM. In addition, Web Figure 2 presents the counterpart of Figure \ref{fig:numeri} when $\rho_2=0.5$. Clearly, a larger CV of cluster sizes leads to a larger correction factor for estimating $\beta_k$ and hence reduces the efficiency, for both LMM and MLMM analyses. However, comparisons between $\bm{\Theta}_{kk}$ and $\theta_k$ imply that MLMM can mildly protect against efficiency loss due to unequal cluster sizes from separate LMM analyses when the inter-subject between-endpoint ICC becomes larger, or the intra-subject ICC is smaller (Web Figure 2).  Under unequal cluster sizes, the intuition underlying the efficient improvement from MLMM over separate LMM analyses is that the FGLS estimator of $\beta_k$ obtained from MLMM additionally depends on the between-endpoint ICCs ($\rho_1^{kk^\prime}$, $\rho_2^{kk^\prime}$) in a complex fashion, whereas the corresponding FGLS estimator of $\beta_k$ obtained from separate LMM only depends on the endpoint-specific ICC ($\rho_0^k$). This is in sharp contrast to the case with equal cluster sizes, where the FGLS estimators of $\beta_k$ obtained from MLMM and separate LMM analyses are identical and are free of any ICCs (Theorem \ref{thm0}). Finally, in Figure \ref{fig2:omnibus}, Web Figures 3 and 4, we numerically explore the relationship between power of the general linear hypothesis test and the intersection-union test with different ICC parameters based on variance expression \eqref{eq:vark}. 
Our limited numerical studies suggest that the findings are consistent with the analytical results derived in Section \ref{s:testequality} and Section \ref{s:testoverall} even when the cluster sizes are unequal. 

\section{Simulation Study}
\label{s:simu}
We evaluate the proposed method in terms of achieving the desired level of power while maintaining the nominal type I error rate through simulation studies. To focus ideas, we consider the intersection-union test to jointly study the treatment effect across $K$ co-primary endpoints, where $K\in\{2,3\}$ is chosen as the most commonly reported values in practice \citep{offen2007multiple}. We assume balanced randomization such that $\overline z = 1/2$ and  $\sigma_z^2=1/4$. For simplicity, we consider $\rho_1^{kk^\prime}= \rho_1$, $\rho_2^{kk^\prime}= \rho_2$ $\forall~k\neq k^\prime$ but set $\bm{\rho}_0 = (\rho_0^1,\ldots,\rho_0^K)^T$ to be a length-$K$ equal-distanced sequence between $\kappa$ and $0.1$. We vary $\kappa \in \{0.01, 0.05\}$ and assume $0.1$ to be the upper bound of $\bm{\rho}_0$ to represent common endpoint-specific ICC values reported in the parallel CRT literature \citep{murray2003methods}. We further specify $\rho_1=\kappa/2$ such that $\rho_1$ is smaller than each element of $\bm{\rho_0}$, and consider $\rho_2 \in \{0.2, 0.5\}$, representing moderate values of the intra-subject ICC. Throughout we fix the marginal endpoint variance $(\sigma_{y1}^2,\ldots,\sigma_{yK}^2)^T=(1,\ldots, K)^T$, and specify the variance component matrices $\bm{\Sigma_\phi}$ and $\bm{\Sigma_e}$ based on the marginal variances and ICCs. To further assess the accuracy of our approximate power procedure under unequal cluster sizes, we consider mean cluster size $\overline{m}=60$, and coefficient of variation of cluster size $\text{CV} \in \{0, 0.2, 0.4, 0.8\}$, representing different degrees of variability in cluster size used in previous simulations \citep{li2021sample}. When $\text{CV}>0$, the cluster sizes $m_i$'s are drawn from a Gamma distribution with shape and scale parameter $1/\text{CV}^2$ and $\overline{m}\text{CV}^2$, rounded to the nearest integer. We vary the true treatment effect parameters $\bm{\beta}=(\eta,0.7)^T$ and $\bm{\beta}=(\eta,(\eta+0.7)/2,0.7)^T$ for $K=2,3$, where $\eta\in\{0.3,0.5\}$. Web Table 1 summarizes the simulation parameters for a quick reference. 

For each of the above parameter combinations, we solve equation \eqref{eq: power2} for $n$ to obtain the required number of clusters to achieve at least $80\%$ power (based on $5\%$ nominal type I error rate), rounded to the nearest even integer above. We ensure that the estimated $n$ is no larger than 30 to resemble typical number of clusters in published parallel CRTs \citep{ivers2011impact}. Given the estimated $n$, we then simulate $K$ continuous co-primary endpoints from MLMM \eqref{eq:model1}, and fit the MLMM to obtain the point and variance estimates for $\bm{\beta}$. We consider the EM approach for estimating the model parameters (details in {Web Appendix A}). To perform the intersection-union test, we set the critical values $c_1=\dots= c_K=t_{\alpha}(n-2K)$, and calculate the empirical power as the proportion of $\mathbb{I}\{\bigcap_{k=1}^K(\zeta_k>c_k)\}=1$ across 1000 simulated CRTs. In each scenario, we compare the empirical power by simulation and the predicted power by formula \eqref{eq: power2} to assess the accuracy of our procedure. Finally, we follow \citet{li2020power} and report the empirical type I error rate as the proportion of false rejections when the CRTs are simulated under $\bm{\beta}=(0, 0.7)^T$ for $K=2$ and $\bm{\beta}=(0,(\eta+0.7)/2,0.7)^T$ for $K=3$.

\begin{table}[htbp]\centering
\caption{Estimated required number of clusters $n$, predicted power $\psi$, empirical power $\overline\psi$ , and type I error rate $e$ with $K=2$, different levels of effect sizes, CV of cluster sizes, ICC values and mean cluster sizes $\overline m$.}\label{tb:k2}
\resizebox{!}{0.65\textwidth}{

\begin{tabular}{crrrrrrrcrrrr}
\toprule
&&& && \multicolumn{2}{c}{$\overline m = 60$} & \phantom{a} & && \multicolumn{2}{c}{$\overline m=80$} \\
\cmidrule{6-7} \cmidrule{11-12} 
Effect Size & CV & $\kappa$ & $\rho_2$ & $n$ & $\psi$ & $\overline\psi$ & $e$ &  & $n$ & $\psi$ & $\overline\psi$ & $e$ \\ 
  \hline
(0.3, 0.7) & 0.0 & 0.01 & 0.20 & 16 & 0.841 & 0.854 & 0.048 &  & 14 & 0.804 & 0.823 & 0.055 \\ 
   &  & 0.01 & 0.50 & 16 & 0.843 & 0.855 & 0.048 &  & 14 & 0.806 & 0.822 & 0.059 \\ 
   &  & 0.05 & 0.20 & 22 & 0.810 & 0.815 & 0.050 &  & 22 & 0.831 & 0.847 & 0.057 \\ 
   &  & 0.05 & 0.50 & 22 & 0.812 & 0.828 & 0.050 &  & 22 & 0.833 & 0.840 & 0.052 \\ 
   & 0.2 & 0.01 & 0.20 & 16 & 0.838 & 0.865 & 0.046 &  & 14 & 0.802 & 0.824 & 0.037 \\ 
   &  & 0.01 & 0.50 & 16 & 0.841 & 0.871 & 0.047 &  & 14 & 0.804 & 0.826 & 0.044 \\ 
   &  & 0.05 & 0.20 & 22 & 0.807 & 0.835 & 0.063 &  & 22 & 0.829 & 0.845 & 0.074 \\ 
   &  & 0.05 & 0.50 & 22 & 0.809 & 0.823 & 0.059 &  & 22 & 0.830 & 0.849 & 0.079 \\ 
   & 0.4 & 0.01 & 0.20 & 16 & 0.832 & 0.844 & 0.039 &  & 16 & 0.856 & 0.865 & 0.059 \\ 
   &  & 0.01 & 0.50 & 16 & 0.834 & 0.847 & 0.042 &  & 16 & 0.858 & 0.870 & 0.061 \\ 
   &  & 0.05 & 0.20 & 24 & 0.836 & 0.855 & 0.058 &  & 22 & 0.822 & 0.841 & 0.046 \\ 
   &  & 0.05 & 0.50 & 24 & 0.837 & 0.865 & 0.058 &  & 22 & 0.823 & 0.833 & 0.050 \\ 
   & 0.8 & 0.01 & 0.20 & 16 & 0.800 & 0.784 & 0.043 &  & 16 & 0.833 & 0.833 & 0.049 \\ 
   &  & 0.01 & 0.50 & 16 & 0.805 & 0.804 & 0.050 &  & 16 & 0.836 & 0.841 & 0.055 \\ 
   &  & 0.05 & 0.20 & 24 & 0.802 & 0.810 & 0.046 &  & 24 & 0.831 & 0.867 & 0.060 \\ 
   &  & 0.05 & 0.50 & 26 & 0.832 & 0.841 & 0.047 &  & 24 & 0.828 & 0.856 & 0.054 \\ 
  (0.5, 0.7) & 0.0 & 0.01 & 0.20 & 14 & 0.814 & 0.854 & 0.041 &  & 14 & 0.825 & 0.839 & 0.055 \\ 
   &  & 0.01 & 0.50 & 14 & 0.814 & 0.857 & 0.047 &  & 14 & 0.825 & 0.839 & 0.059 \\ 
   &  & 0.05 & 0.20 & 16 & 0.853 & 0.870 & 0.054 &  & 14 & 0.809 & 0.826 & 0.062 \\ 
   &  & 0.05 & 0.50 & 16 & 0.854 & 0.874 & 0.055 &  & 14 & 0.810 & 0.829 & 0.066 \\ 
   & 0.2 & 0.01 & 0.20 & 14 & 0.813 & 0.838 & 0.037 &  & 14 & 0.824 & 0.846 & 0.037 \\ 
   &  & 0.01 & 0.50 & 14 & 0.813 & 0.834 & 0.035 &  & 14 & 0.824 & 0.843 & 0.044 \\ 
   &  & 0.05 & 0.20 & 16 & 0.851 & 0.875 & 0.061 &  & 14 & 0.808 & 0.832 & 0.052 \\ 
   &  & 0.05 & 0.50 & 16 & 0.852 & 0.882 & 0.061 &  & 14 & 0.809 & 0.838 & 0.060 \\ 
   & 0.4 & 0.01 & 0.20 & 14 & 0.808 & 0.829 & 0.045 &  & 14 & 0.821 & 0.863 & 0.045 \\ 
   &  & 0.01 & 0.50 & 14 & 0.809 & 0.826 & 0.040 &  & 14 & 0.821 & 0.867 & 0.051 \\ 
   &  & 0.05 & 0.20 & 16 & 0.846 & 0.862 & 0.050 &  & 14 & 0.803 & 0.840 & 0.063 \\ 
   &  & 0.05 & 0.50 & 16 & 0.847 & 0.859 & 0.044 &  & 14 & 0.805 & 0.845 & 0.061 \\ 
   & 0.8 & 0.01 & 0.20 & 16 & 0.842 & 0.841 & 0.043 &  & 14 & 0.806 & 0.854 & 0.059 \\ 
   &  & 0.01 & 0.50 & 16 & 0.847 & 0.851 & 0.050 &  & 14 & 0.810 & 0.851 & 0.071 \\ 
   &  & 0.05 & 0.20 & 16 & 0.822 & 0.817 & 0.054 &  & 16 & 0.842 & 0.857 & 0.051 \\ 
   &  & 0.05 & 0.50 & 16 & 0.825 & 0.820 & 0.057 &  & 16 & 0.845 & 0.853 & 0.057 \\ 
\bottomrule
\end{tabular}
}
\end{table}

Table \ref{tb:k2} presents the estimated required number of clusters $n$, predicted power, empirical power, and type I error rate with $K=2$ co-primary endpoints and $\overline m\in \{60, 80\}$.
Consistent with the exploration in Section \ref{s:cv}, the estimated number of clusters $n$ increases mildly when the CV of cluster sizes increases. Overall, the empirical powers of the Wald tests are in reasonable agreement with the predicted powers by the proposed formula, and the empirical type I error rates are generally close to the nominal level. Similar trends are also observed in Web Table 2 with $K=3$ co-primary endpoints.  Web Tables 3 and 4 summarize the expected standard errors of the empirical power, which are all around $1\%$ and thus fairly small \citep{morris2019using}. Finally, Web Tables 5 to 10 summarize the bias in estimating the variance component parameters in $\bm{\Sigma_{\phi}}$, $\bm{\Sigma_e}$ with $K=2$ and $K=3$ co-primary endpoints. Across all simulation scenarios with no more than $30$ clusters, the EM approach leads to relatively small bias in estimating the variance parameters, suggesting no evidence of non-identifiability.

\section{Application to the Kerala Diabetes Prevention Program Trial}
\label{s:application}
We illustrate the proposed method using the Kerala Diabetes Prevention Program (K-DPP) study \citep{thankappan2018peer}, which is a parallel CRT aimed to evaluate the efficacy of a peer-support lifestyle intervention in preventing type 2 diabetes among high-risk individuals. A total of 60 polling areas (clusters) were randomized in a 1:1 ratio to either participate in the peer-support program (intervention) or simply receive the education booklet (usual care). The study included two secondary clinical endpoints: 
change in systolic and diastolic blood pressure measured from baseline to 24 months. To illustrate our new methodology, we consider a scenario where the investigators are interested in planning a CRT to study the effect of the K-DPP intervention on two continuous co-primary endpoints: change in systolic and diastolic blood pressure ($K=2$). 
We consider the omnibus test and the intersection-union test, and determine the number of clusters required to achieve 80\% power at the 5\% significance level when each one of these tests are of primary interest. In the context of the K-DPP study, rejecting the null with the omnibus test means that the peer-support lifestyle program has an effect on at least one of the systolic and diastolic blood pressure outcomes, whereas rejecting the null with the intersection-union test means that the peer-support lifestyle program has an effect on both outcomes. To proceed, we estimate the design parameters from the K-DPP study. The mean cluster size is estimated from the study as $\overline{m}=17$ and the CV of cluster size is $0.19$. We fit the MLMM using the EM algorithm and obtain the variance matrices as $\bm{\Sigma_\phi}=\left(\begin{array}{@{}cc@{}}
                        8.3 & 9.1 \\
                        9.1 & 11.2
                         \end{array}\right)$ and $\bm{\Sigma_e}=\left(\begin{array}{@{}cc@{}}
                        170.0 & 94.2 \\
                        94.2 & 84.8
                         \end{array}\right)$.
These values correspond to marginal variances $\bm{\sigma}^2_y=(178.4, 96.0)$ and ICC values $(\rho_0^1, \rho_0^2, \rho_1^{12}, \rho_2^{12})=( 0.05, 0.12, 0.07, 0.79)$.  For the omnibus test, solving equation \eqref{eq: power1} with variance \eqref{eq:vark} suggests that $n=48$ clusters are required to detect effect sizes $(\beta_1, \beta_2)= 0.3\times \bm{\sigma}_y$ with 80\% power. Furthermore, for the intersection-union test, solving equation \eqref{eq: power2} with variance \eqref{eq:vark} suggests that $n=50$ clusters are needed to detect effect sizes $(\beta_1, \beta_2)= 0.3\times \bm{\sigma}_y$ with 80\% power.

\begin{figure}[htbp]
\centering
\includegraphics[width = 1\textwidth]{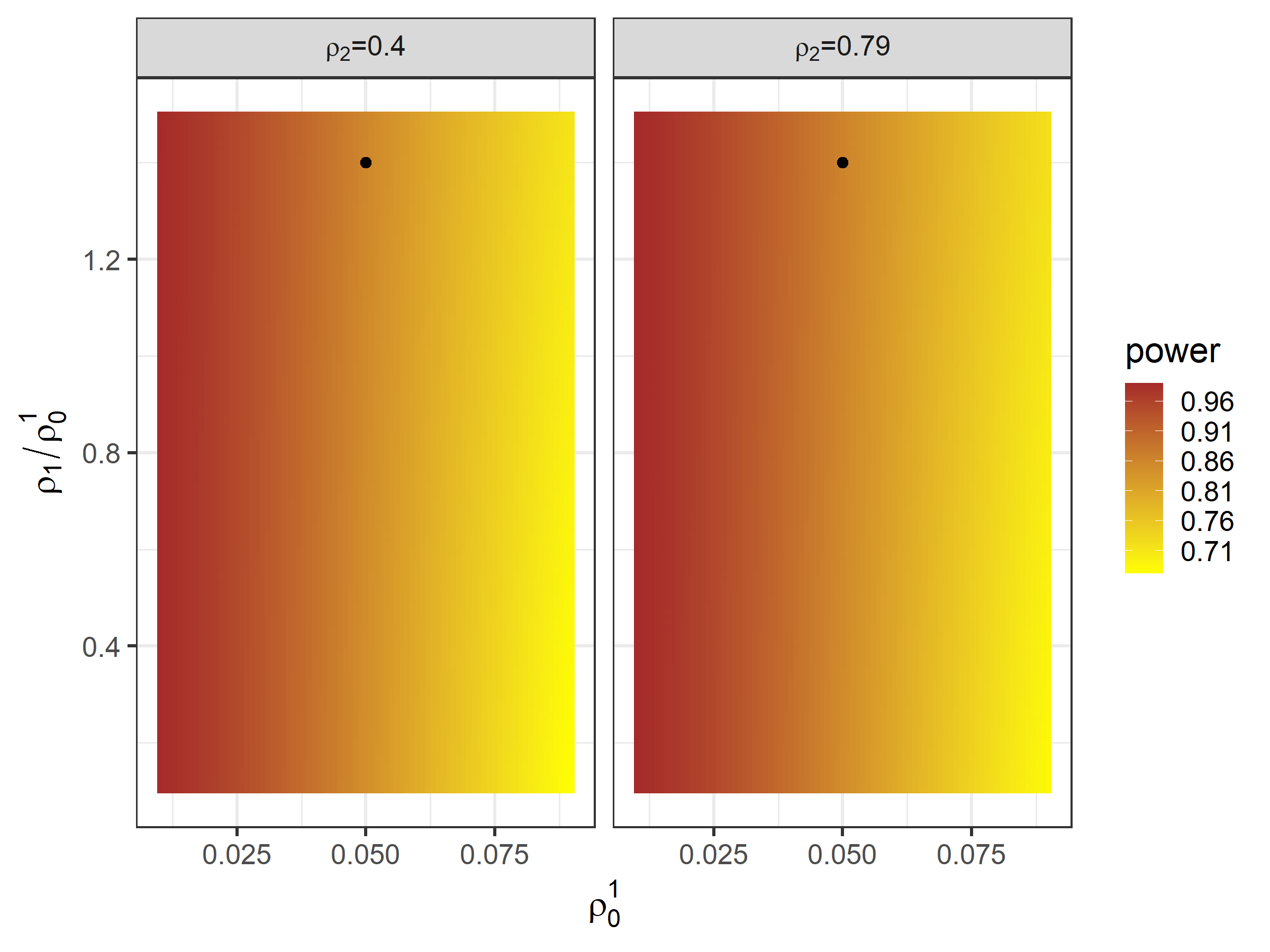}
\caption{Predicted power for the intersection-union test with $n=60$ clusters with varying ICC values as additional sensitivity analysis. The predicted power corresponding to the ICC values estimated from the K-DPP trial is highlighted with a solid black dot.}
\label{f:heatmap}
\end{figure}

Assuming that the study can feasibly recruit up to $n=60$ clusters (the actual number of clusters in the K-DPP study), we further investigate the range of power predictions based on our formulas when the ICC values deviate from the above design assumptions. 
In this evaluation, we vary the ICC for change in systolic blood pressure $\rho_0^1\in[0.01,0.09]$, and fix $\rho_0^2=2.4\times \rho_0^1$ based on the proportionality relationship in the estimates in the K-DPP study. Figure \ref{f:heatmap} presents a power contour for the intersection-union test when the inter-subject between-endpoint ICC $\rho_1^{12}/\rho_0^1\in\times [0.1,1.5]$, intra-subject ICC $\rho_2^{12} \in \{0.4, 0.79 \}$ for the effect size $(\beta_1, \beta_2)= 0.3\times \bm{\sigma}_y$. The predicted power ranges between $[0.67, 0.99]$, and it is evident that power decreases with larger values of $\rho_0^1$ and smaller values of $\rho_2^{12}$. Web Figure 5 present the corresponding power contour for the omnibus test. Across the range of ICC values we considered, the predicted power of the omnibus test are within the range $[0.76, 1.00]$. The figure indicates that the larger between-endpoint ICCs lead to smaller power of the omnibus test, matching the theoretical prediction from Theorem \ref{thm1}, even when the cluster sizes are mildly variable. Finally, we provide an illustrative example of power calculation based on the test for treatment effect homogeneity in Web Appendix F.

\section{Discussion}
\label{s:discuss}
In this article, we have developed a new analytical approach for power analysis of CRTs with continuous co-primary endpoints, addressing one of the pressing challenges in many current pragmatic clinical trials with multivariate endpoints \citep{Taljaard2021}. Specifically, we describe a MLMM to account for three different types of ICCs within each cluster: the endpoint-specific ICC, the inter-subject between-endpoint ICC and the intra-subject ICC. In addition, we derive the joint distribution of the vector of treatment effect estimators based on the feasible generalized least squares approach, and elucidate the impact of different ICCs on power for three types of tests that can be considered for analyzing multivariate endpoints. We show that the usual implications of the endpoint-specific ICC values hold in the multivariate setting, namely higher ICC values are associated with larger required sample size. For the inter-subject between-endpoint ICCs and the intra-subject ICCs, it is difficult to predict generally, but under conditions of block-exchangeability, we show that their implications differ, depending on whether the omnibus test or intersection-union test is used. For the omnibus test, higher values for these ICCs lead to a larger required sample size, whereas for the intersection-union test, larger values lead to a smaller required sample size. We also show that when cluster sizes vary, using a multivariate approach has advantages in that larger values of inter-subject between endpoint ICCs can possibly protect against efficiency loss due to cluster size variability. Finally, we extend our approach to accommodate unequal cluster sizes, where the power formula further depends on the CV of the cluster size distribution. Our simulation study suggests that the power formula is accurate even when there is a limited number of clusters as well as small to large degree of cluster size variation, which encompass frequent scenarios seen in CRT applications \citep{ivers2011impact} 

 With multivariate co-primary outcomes, the intersection-union test is a simple and practical approach when the interest lies in detecting treatment effect signals across all endpoints. To operationalize this test for study design, we considered a multivariate $t$-distribution with degrees of freedom, $n-2K$. While in our simulation studies with $n\geq 14$ and $K\in\{2,3\}$, the intersection-union test has demonstrated nominal type I error rate and adequate empirical power, this test is likely conservative when either the number of clusters further decreases or the number of endpoints further increases. For example, in a CRT with $8$ clusters and $K=3$ endpoints, the critical value $t_{\alpha}(n-2K=2)$ will be substantially larger than the corresponding normal critical value, and therefore the intersection-union test may frequently fail to reject the null. This is also the situation where the empirical power of this test may be low. The implication of this observation for study planning with co-primary outcomes is that the number of clusters should be at least a handful to support $K\in\{2,3\}$ co-primary endpoints. With an even larger number of endpoints, additional clusters will be necessary to ensure sufficient degrees of freedom for the intersection-union test, and it would be useful to develop a rule of thumb in future research.

A different approach for power analysis of CRTs is based on the marginal model coupled with GEE for parameter estimation \citep{preisser2003integrated}. For instance, with binary co-primary endpoints, \citet{li2020power} developed the analytical variance expression of the GEE treatment effect estimators assuming equal cluster sizes and an independence working correlation structure. It is possible to extend their approach to accommodate continuous co-primary endpoints as an alternative to our proposed method. However, previous work has shown that a GEE with an independence working correlation structure can result in an inflated sample size compared to an efficient GEE with correct working correlation model even with a single primary endpoint \citep{li2021sample}. It would be interesting to quantify the efficiency gain by MLMM versus independence GEE with co-primary endpoints. Second, we have assumed all co-primary endpoints are continuous, whereas in certain applications there can be a mix of continuous and binary co-primary endpoints. It would be worthwhile to further develop our method to accommodate co-primary endpoints with mixed types. Finally, co-primary endpoints can also arise in multiple-period CRTs, which requires consideration of even more complex correlation structures. We plan to pursue an extension of our methods to multiple-period CRTs (e.g., stepped wedge CRTs) in our future work.

\section*{Acknowledgements}
Research in this article was supported by a Patient-Centered Outcomes Research Institute Award\textsuperscript{\textregistered} (PCORI\textsuperscript{\textregistered} Award ME-2020C3-21072), as well as by the National Institute of Aging (NIA) of the National Institutes of Health (NIH) under Award Number U54AG063546, which funds NIA Imbedded Pragmatic Alzheimer’s Disease and AD-Related Dementias Clinical Trials Collaboratory (NIA IMPACT Collaboratory). The statements presented are solely the responsibility of the authors and do not necessarily represent the views of PCORI\textsuperscript{\textregistered}, its Board of Governors or Methodology Committee, or the National Institutes of Health. The authors thank the Associate Editor and two anonymous referees for their constructive comments, which greatly helped improve the exposition of our work.

\section*{Data Availability Statement}
Data used in this paper as an illustrative example are publicly available from the figshare database \citep{Thirunavukkarasu2017} at \url{https://doi.org/10.6084/m9.figshare.5661610}.

\vspace{-0.2in}
\bibliographystyle{biom}
\bibliography{Coprimary}

\vspace{-0.2in}
\section*{Supporting Information}
Web Appendix,  R code for fitting the EM algorithm for the multivariate linear mixed model, functions for power and sample size determination, and code used in the numerical study and data application are available at  \url{https://github.com/siyunyang/coprimary_CRT}.

\label{lastpage}
\end{document}